\documentstyle[preprint,eqsecnum,psfig,aps]{revtex}
\tighten

\begin{document}
\draft
\preprint{} 
\title{CMB Anisotropy Due to Tangled magnetic fields in re-ionized models
\\}
\author{T. R. Seshadri$^1$ and Kandaswamy Subramanian$^2$\\
$^1$ {\small Department of Physics and Astrophysics, University of Delhi,\\
Delhi 110007, India}\\
$^2${\small Inter-University Centre for Astronomy and Astrophysics, Post Bag 4, Ganeshkhind,\\ 
Pune 411 007, India.} \\}
\maketitle

\tightenlines
\def\btt#1{{\tt$\backslash$#1}}
\def\BibTeX{\rm B{\sc ib}\TeX}
\draft
\maketitle
\begin{abstract}
Primordial tangled cosmological Magnetic Fields source 
rotational velocity perturbations of the baryon fluid, even 
in the post-recombination universe. 
These vortical modes inturn leave a characteristic
imprint on the temperature anisotropy of the Cosmic 
Microwave Background (CMB),
if the CMB photons can be re-scatterred after recombination. 
Observations from WMAP indicate that the Universe underwent a 
relatively early re-ionization ($z_{ri} \sim 15$), which
does indeed lead to a significant optical depth for
re-scattering of CMB photons
after the re-ionization epoch. 
We compute the resulting additional temperature anisotropies, induced by 
primordial magnetic fields in the post-recombination universe.
We show that in models with early re-ionization, a nearly 
scale-invariant spectrum of tangled magnetic fields which 
redshift to a present value of $B_0 \sim 3 \times 10^{-9}$ Gauss,
produce vector mode perturbations which in turn induce 
additional temperature anisotropy of about $0.3$ to $0.4$ $ \mu$K over 
very small angular scales, with $ l$ upto $\sim 10000 $ or so. 
\end{abstract}


\section{Introduction\protect\\}
\label{introduction}
There are two possible processes which might explain the origin 
of large scale cosmic magnetic fields. 
Both have, however, potential difficulties.
One possibility is that some high-energy process in the early universe 
(like inflation or a cosmological phase transition) generated 
primordial magnetic fields which manifest today as galactic or 
cluster fields  \cite{primhyp}. The problem here is that
this involves speculative physics and there is as yet no compelling
mechanism to generate fields of the required strength \cite{grasso,widrow2003,giovannini}.
Alternatively, seed magnetic fields can get amplified by a
large scale dynamo, to produce fields as observed today, 
coherent over galactic or cluster scales \cite{dynam,ab_ks_phys_rep}.
There are, however, constraints from helicity conservation and/or suppression
of lagrangian chaos, due to which the efficacy of this process
is unclear (see for example \cite{ab_ks_phys_rep}
and references therein).
The effects of weak primordial magnetic fields 
(whose strength today is of order
$10^{-9}$ Gauss), and which
are tangled on galactic scales can affect
galaxy formation 
\cite{rees_reinhardt,wasserman,kim_olinto_rosner,SB98a,sethi03}
Hence, it is of considerable interest, to find different ways 
of limiting or detecting such primordial fields 
(see \cite{grasso,widrow2003,kron} for reviews).

In several earlier papers the consequences of tangled primordial 
cosmological magnetic fields on the observable signatures 
on the CMBR anisotropy and polarization have been investigated 
in a cosmological scenarios with no re-ionization 
\cite{trs_ks,ksjdb,ks_trs_jdb,mkk02,dff,SB98}.
A detailed numerical investigation of CMB signals
due to tangled magnetic fields has also been undertaken in
 \cite{lewis}.
The First year results from WMAP satellite observations however
indicate that the universe could have undergone an early stage 
of re-ionization \cite{kovac,kogut}, with 
a large optical depth $\kappa \sim 0.17$ to the re-scattering
of CMB photons.
In this paper we follow up our earlier work by investigating the
consequences of a tangled primordial magnetic field, for CMB
anisotropies, in a scenario with early re-ionization.
In particular we focus on the rotational perturbations, 
that are produced by primordial fields in the post-recombination 
universe, and the additional CMB anisotropy signals 
that they induce in a universe with early re-ionization. 
Note that compressive velocity modes
could also induce CMB anisotropies; however they also have
non magnetic sources and more importantly suffer
larger cancellation effects due to the thickness of the
last scattering surface around the re-ionization epoch.
Hence, our focus is only on rotational modes here.

In section II the general formulation of the problem
and the parameters of re-ionization
relevant for our calculation are derived.  Semi-analytic 
estimates of the additional CMB signals are made in section III. 
In Section IV we present numerical calculations and 
in Section V we discuss our results. 

\section{General formulation\protect\\}
\label{formulation}
The equations for the evolution of temperature anisotropy 
for scalar, vector and tensor modes have been derived by 
Hu and White \cite{huwhit} (hereafter referred to as HW97). 
We concentrate here on the
additional contributions which arise in a re-ionized universe, 
due to the vector modes induced by inhomogeneous magnetic fields. 
From equation (74) and (56) of HW97, the angular power-spectrum of CMB 
anisotropy corresponding to vector modes is given by,
\begin{equation}
{C_l}=
4 \pi {\int}dk~\frac{k^2}{2{\pi^2}}\frac{l(l+1)}{2}
\langle \mid \int_0^{\tau_0}d{\tau}~g(\tau_0,\tau)V(k,\tau)
\frac{j_l(k(\tau_0-\tau))}
{k(\tau_0 - \tau)}  \mid^2  \rangle   \label{Cl}
\end{equation}
Here, $V(k,\tau )$ is the magnitude of the vorticity, generated by
tangled primordial magnetic fields, in Fourier space.
Also, $k$ is the co-moving wave number, $\tau $ is conformal time, 
$\tau _{0}$ its present value, and $j_{l}(z)$ is the spherical Bessel function 
of order $l$. We have ignored a small polarization correction to the source term
and also a metric perturbation term which are in general sub-dominant
(cf. \cite{trs_ks,mkk02}). The
'visibility function', $g(\tau _{0},\tau ),$ is given by,
\begin{equation}
g(\tau_0,\tau)=  n_e(\tau) \sigma_Ta(\tau)
\exp \left [-\int^{\tau_0}_{\tau}
  n_e(\tau') \sigma_Ta(\tau')d{\tau'} \right],   \label{g}
\end{equation}
where $g(\tau_0,\tau)d\tau$ is the probability that
a photon that reaches us at epoch $\tau _{0}$ was last scattered between the
epochs $(\tau,\tau + d\tau) $. We assume a flat universe throughout, 
with a total  
matter density parameter $\Omega _{m}$ and a non-zero cosmological 
constant density parameter $\Omega_{\Lambda }$ today. 

The exact form of the visibility function is determined by 
the ionization history of the universe. 
For standard recombination $g(\tau_0,\tau)$ has only one peak
around the epoch of recombination. However re-ionization
can modify this further by making $g$ significant after
the universe has been re-ionized, the exact modification
depending on the complex ionization history of the baryons.
The modifications of the visibility function will 
show up in the power spectrum of the anisotropy. In this paper we will not be 
focusing our attention on the physical mechanism responsible for re-ionization.
We will, however, assume that whatever be the source, it results in a sharp 
transition to a re-ionized situation. We assume this transition to be a
step function at a redshift $z_{ri}$ corresponding to a 
conformal time $\tau_{ri}$. The value of $\tau_{ri}$ is fixed in this 
simple model by the optical depth indicated by the WMAP results. 
The visibility function will then have 
two peaks, one at $\tau_{rec}$ and the other at $\tau_{ri}$. 

The free electron number density in the 
era after standard recombination is modelled as,
\begin{equation}
n_e=\frac{n_{b0}}{a^3}{\Theta(\tau-\tau_{ri})}  \label{ne_theta_f}
\end{equation}
where $n_{b0} = \Omega_{b}3H_0^2/(8\pi G m_b)$ with
$\Omega_{b}$ being the density parameter of the Baryons.
(We have neglected the small residual electron density
after recombination).
The optical depth is given by
\begin{equation}
\kappa=c{\sigma_T} \int_t^{t_0} n_e(t')~dt' = 
c{\sigma_T}\int_\tau^{\tau_0}n_e(\tau') a(\tau')~d\tau' .
\end{equation}
Here the proper time $t$ and conformal time $\tau$ are related
by using $dt = ad\tau$ and $(t_0,\tau_0)$ correspond to the present epoch.
The integration over $t$ can be converted into 
an integration over redshift by the substitution,
$dt=da/(Ha) =-(H(z)(1+z))^{-1}~dz$. Here $H(z)$ is 
the Hubble expansion rate and is given by
\begin{equation}
H(z)=H_0[\Omega_{\Lambda}+\Omega_{m}(1+z)^3]^{1/2}.
\end{equation}
The optical depth up to re-ionization can now be expressed as,
\begin{equation}
\kappa_{ri}=c{\sigma_T}\frac{{\Omega_{b}{H_0}}}{8{\pi}Gm_b}
\frac{\sqrt \Omega_{\Lambda}}
{\Omega_{m}}\int_{z=0}^{z=z_{ri}}dy~\frac{1}{\sqrt{1+y}}
\end{equation}
where $y$ is defined as $y=(1+z)^3(\Omega_{m}/\Omega_{\Lambda})$.
On integration, this gives,
\begin{equation}
\kappa_{ri}=\frac{cH_0{\Omega_{b}}}{4{\pi}Gm_b}
\frac{\sqrt{\Omega_{\Lambda}}}{\Omega_{m}}
\sigma_T\left[~~
\sqrt{1+(1+z_{ri})^3\frac{\Omega_{m}}{\Omega_{\Lambda}}}
~~~-~~~\sqrt{1+\frac{{\Omega_{m}}}{{\Omega_{\Lambda}}}}~~
\right]
\end{equation}

Adopting, $\kappa_{ri} =0.17$, $h=0.71$, 
$\Omega_{b}=0.044$, $\Omega_{m}=0.27$ and $\Omega_{V}=0.73$,
we get, $z_{ri}=14.57$. (This value is close to the one obtained from 
equation 24.79 given in Peebles
\cite{peebles_PPC}. We have neglected a small correction due to helium
fraction. This, however, does not significantly affect the results.)
The conformal time of re-ionization is given by,
\begin{equation}
\tau_{ri}=\frac{2cH_0^{-1}}{\sqrt{\Omega_{m}}}\left( \sqrt{a_{ri}+a_{eq}}-
\sqrt{a_{eq}}\right)
\end{equation}
where, $a_{ri}^{-1}=1+z_{ri}$ and $a_{eq}^{-1}=1+z_{eq}$ 
specifies the scale factor at matter-radiation equality.
The current epoch is given by
\begin{equation}
\tau_0=\frac{2cH_0^{-1}}{\sqrt{\Omega_{m}}}\left[1-0.0841~~ln(\Omega_{m})
\right]\left(\sqrt{1+a_{eq}}-\sqrt{a_{eq}}\right)  
\end{equation}
With $2cH_0^{-1}= 6000~h^{-1}$Mpc, $z_{eq}=3233$ and $z_{ri}=14.57$, we get 
$\tau_0=12595h^{-1}$Mpc, $\tau_{ri}=2741h^{-1}$Mpc and 
$\tau_0-\tau_{ri}=9854h^{-1}$Mpc.
Using the functional form of the number density of the electrons as given in 
equation (\ref{ne_theta_f}) and the $\tau_{ri}$ determined above,
 we can calculate the form of the visibility function.  
We will use this form for the visibility function in the numerical 
computation in section \ref{numres}. However, to begin with, we  
approximate the visibility function $g_2(\tau_0,\tau)$ after recombination, 
to be a truncated exponential, 
and estimate the CMB anisotropy in a semi-analytic manner.

Specifically, we adopt,
\begin{equation}
g_2(\tau_0,\tau)=\frac{N_2}{\alpha}e^{-\frac{\tau-\tau_{ri}}{\alpha}}
{\Theta(\tau-\tau_{ri})}
\end{equation}
Here the Heavyside $\theta(x)$ function, is zero for $x < 0$ and $1$
for $x > 0$. It takes account of the fact that before re-ionization,
$n_e$ is negligible. Further, $N_2$ is a normalization consant and 
$\alpha$ gives the spread of the 
exponential. By appropriately choosing $\alpha$, 
we can set the width of the reionized last scattering surface.
Also note that $g(\tau_0,\tau)$ has the interpretation of probability; so
its integral over $\tau$ from $\tau=0$ to $\tau=\tau_0$ should be normalized 
to unity.  This determines the normalization factor $N_2$.
For a sufficiently early epoch of re-ionization, we 
generally have $(\tau_0 -\tau_{ri})/\alpha \gg 1$.
In this case, the condition that the integral  of 
$g(\tau_0,\tau)$ over $\tau$ should be unity implies 
$N_2 + e^{-\kappa_{ri}} =1$,
or $N_2 = 1 - \exp-(\kappa_{ri})$. So $N_2$ measures the probability
of at least one scattering between $\tau_0$ and $\tau_{ri}$,
due to the re-ionization. For small $\kappa_{ri} \ll 1$, we have
$N_2 \sim \kappa_{ri}$

The constant $\alpha$ is determined by the (conformal) time, say $\tau_m$, 
after which $g_2$ drops to $1/e$ times its peak value (at $\tau_{ri}$). Thus if 
\begin{equation}
\frac{g_2(\tau_0,\tau_m)}{g_2(\tau_0,\tau_{ri})}=\frac{1}{e}
\end{equation}
then, $\alpha=\tau_m-\tau_{ri}$. To determine $\tau_m$, we use the exact form 
of $g_2$ and calculate the epoch when $[g(\tau_0,\tau_{ri})/
g(\tau_0,\tau_{m})] =e$. Using equation 
(\ref{ne_theta_f}) and the expression for the 
visibility function in equation (\ref{g}), we get,
\begin{equation}
\frac{   g(\tau_0,\tau_{ri})        }{  g(\tau_0,\tau_{m})  }
=e =\frac{n_e(\tau_{ri}) a(\tau_{ri})}{n_e(\tau_{m}) a(\tau_{m})} \
\exp\left[-\int_{\tau_{ri}}^{\tau_m} n_e(\tau'') \sigma_Ta(\tau'')d{\tau''}
\right]
\end{equation}
In the interval between $\tau_{ri}$ and $\tau_m$, the universe is in general 
matter dominated. Hence $a(\tau) \propto \tau^2$. Also for $\tau>\tau_{ri}$, we 
have, $n_e \propto a^{-3}$.
Hence, we can simplify the above equation to
\begin{equation}
\frac{\tau_m^4}{\tau_{ri}^4}~
\exp \left[-c{\sigma_T}\frac{3{\Omega_{b0}{H_0}^2}}
{8{\pi}Gm_b}\int_{\tau_{ri}}^{\tau_m}\frac{d\tau}{a^2} \right]=e
\end{equation}
Substituting for the scale factor as $a=\tau^2/\tau_0^2$,
we get,
\begin{equation}
4~ln\left(\frac{\tau_m}{\tau_{ri}}\right)-\frac{{\Omega_{m0}}H_0^2\sigma_
Tc\tau_{ri}(1+z_{ri})^2}{8{\pi}Gm_b}\left(1-\frac{\tau_{ri}^3}
{\tau_m^3}\right)=1
\end{equation}
This gives $\tau_m=3519h^{-1}$Mpc for $\tau_{ri}=2741h^{-1}$Mpc giving the 
value of $\alpha$ as $778h^{-1}$Mpc. 

\section{Results of semi-analytic approximation\protect\\}

Consider first the limiting case when
$j_l$ is more sharply peaked than $g_2$. 
This implies the limit $l\gg (\tau_0-\tau_{ri})/\alpha$. 
On using the values of $\tau_0$, $\tau_{ri}$ and $\alpha$ determined
above this limit translates to $l \gg 12$.
We will also assume that the source 
term $V$ varies slower than either the variation of $g_2$ or 
that of $j_l$. These approximations will help in understanding the effect of the model 
with re-ionization in a semi-analytic manner. 
Results of numerical calculations which do not make
these approximations are given in the section \ref{numres}.

Since we wish to focus on the effects of re-ionization we 
take the visibility function to be $g_2$ in the integral appearing in 
equation (\ref{Cl}), which then becomes, 
\begin{equation}
\int_0^{\tau_0}d{\tau}~
g_2(\tau_0,\tau)V(k,\tau)\frac{j_l(k(\tau_0-\tau))}{k(\tau_0-\tau)}. 
\end{equation}
The function, $j_l$ peaks at $l \sim k(\tau_0-\tau)$ or $\tau \sim \tau_0-l/k$. 
So in the above integral we evaluate $V(k,\tau)$ and 
$g_2(\tau_0,\tau)$ at $\tau = \tau_0-l/k$ and replace 
$k(\tau_0-\tau_{ri})$ by $l$ and move 
them out of the integral. We also use the identity,
\begin{equation}
\int_0^{\infty}j_l(x)~dx~=~\sqrt{\frac{\pi}{2l}}.
\end{equation}
Substituting the final result in the expression for 
$C_l$ in equation (\ref{Cl}), we get,
\begin{eqnarray}
C_l&=&\frac{N_2^2}{\alpha^2}\frac{l(l+1)}{2l^3}
\int_0^\infty
dk~k^2 \langle ~\mid
V(k,\tau_0-l/k) \mid^2~\rangle
e^{-2(\tau_0-\tau_{ri}-l/k)/{\alpha}}
{\Theta(\tau_0-\tau_{ri}-l/k)}  \\
&=&\frac{N_2^2}{2{\alpha^2}l}\int_0^{(\tau_0-\tau_{ri})} dx~\frac{k^4}{l}
\langle ~\mid
V(k,\tau_0-l/k) \mid^2~\rangle e^{-(2x/\alpha)} \Theta (x)
\end{eqnarray}
where, we have substituted $x=(\tau_0-\tau_{ri}-\frac{l}{k})$.
The exponential term peaks at $x=0$. 
So we will evaluate the rest of the integrand at $x \to 0$ 
or $k \to l/(\tau_0-\tau_{ri})$ and move it out of the integrand.
The integration of the exponential factor then gives,
\begin{eqnarray}
C_l&=&\frac{N_2^2 2{\pi^2}}{4{\alpha}l^2}
\left( \frac{\tau_0-\tau_{ri}}{l}\right) 
\left[ \frac{k^3
\langle ~\mid
V(k,\tau_0-l/k) \mid^2~\rangle
}{2{\pi^2}}\right]_{k=l/(\tau_0-\tau_{ri})}  \\
\frac{l(l+1)}{2\pi}C_l&=&\frac{N_2^2 {\pi}}{4{\alpha}}
\left( \frac{\tau_0-\tau_{ri}}{l}
\right) \Delta^2_V(k,\tau_{ri})\vert_{k=l/(\tau_0-\tau_{ri})}
\label{ll+1Cl}
\end{eqnarray}
for $l\gg 1$ and $\tau_0-\tau_{ri} \gg \alpha$.
Here we have defined the power spectrum associated with
rotational velocity perturbations, 
\begin{equation}
{\Delta^2_V}(k,\tau)=\frac{k^3\langle~\mid V(k,\tau)\mid^2\rangle}
{2\pi^2} .
\end{equation}

We assume that the magnetic field which induces vortical perturbations 
is initially a Gaussian random field.
On large enough scales, the induced velocity is generally so small
that it does not lead to any appreciable distortion of the 
initial field \cite{SB98a,jko,SS04}. So,
the magnetic field simply redshifts away as ${\bf B}(%
{\bf x},t)={\bf B}_{0}({\bf x})/a^{2}$. The Lorentz force associated with
the tangled field is then ${\bf F}_{L}=({\bf \nabla }\times {\bf B}%
_{0})\times {\bf B}_{0}/(4\pi a^{5})$, which pushes the fluid and creates
rotational velocity perturbations. These can be estimated as in \cite{wasserman}
by using the Euler equation for the baryons and we give a detailed
derivation in Appendix A.

Further, in order to compute the ensemble average 
${\Delta^2_V}$ and hence the $C_{l}$s, 
we need the magnetic spectrum $M(k)$.  This is defined using
$<b_{i}({\bf k})b_{j}({\bf q}%
)>=\delta _{{\bf k},{\bf q}}P_{ij}({\bf k})M(k)$, where $\delta _{{\bf k},%
{\bf q}}$ is the Kronecker delta which is non-zero only for ${\bf k}={\bf q}$%
. Here ${\bf b}({\bf k})$ is the Fourier transform of ${\bf B}_0$, the
present day value of the tangled magnetic field. 
Also $P_{ij}({\bf k}) = (\delta_{ij} -k_ik_j/k^2)$ is the
projection operator which ensures that magnetic field
has zero divergence. 
This gives \\
$<{\bf B}_{0}^{2}>=2\int (dk/k)\Delta _{b}^{2}(k)$, where $%
\Delta _{b}^{2}(k)=k^{3}M(k)/(2\pi ^{2})$ is the power per logarithmic
interval in $k$ space residing in magnetic tangles, and we replace the
summation over $k$ space by an integration. 
It is convenient to define a dimensionless spectrum, $m(k)=\Delta
_{b}^{2}(k)/(B_{0}^{2}/2)$, where $B_{0}$ is a fiducial constant magnetic
field. The baryonic Alfv\'{e}n velocity, $V_{A}$, for this fiducial field is,
\begin{equation}
V_{A}={\frac{B_{0}}{(4\pi \rho_{b0})^{1/2}}}\approx 1.5 \times
10^{-5}B_{-9},  
\label{alfvel}
\end{equation}
where $\rho_{b0}$ is the present day Baryon density, and
$B_{-9} \equiv (B_{0}/10^{-9}{\rm Gauss})$. 

We will also consider
as in \cite{ks_trs_jdb}, power-law magnetic spectra, $M(k)=Ak^{n}$. 
We will cut-off this spectra at the scale where the perturbations
are no longer linear, say at $k=k_{N}$. We expect $k_N$ to be
of order galactic scales, or $k_N \sim (f h Mpc)^{-1}$ with
$f \sim 1$, for the range of redsifts
which make a non-zero contribution to the visibility function
and for the field strengths that we consider \cite{SB98a,SS04}.
We fix $A$ by demanding
that the field smoothed over a scale, $k_{N}$, (using a sharp $k$-space
filter) is $B_{0}$, giving a dimensionless spectrum for $n>-3$ of
\begin{equation}
m(k)=(n+3)(k/k_{N})^{3+n}.
\label{powspec}
\end{equation}

Assuming such a spectrum we have from Appendix A,
\begin{equation}
\frac{k^3\langle~\mid V(k,\tau)\mid^2\rangle}{2\pi^2}=\left(  \frac{2c}{H_0}  
\right)^2 \left(\frac{1+z}{\Omega_{m0}}\right) \left[  \frac{kV_{A}^2}{\sqrt{8}} 
I(k)\right]^2 . \label{velocity-power-spectrum-b}
\end{equation}
Here $I(k)$ is a mode coupling integral which has been
worked out in detail in \cite{SB98,trs_ks}. 
In particular for the case $n<-3/2$ which we consider below it is given by,
\begin{equation}
I^2(k)=\frac{8}{3}(n+3)(\frac{k}{k_N})^{6+2n}.    
\label{I2-b}
\end{equation}
Using equation (\ref{I2-b}) in equation 
(\ref{velocity-power-spectrum-b}) and substituting the resulting expression for 
velocity power-spectrum in equation (\ref{ll+1Cl}) we get,
\begin{equation}
\frac{l(l+1)}{2\pi}C_l=\frac{{\pi}N_2^2}{4}\left( \frac{2c}{H_0} \right)^2~
\frac{1+z_{ri}}{8\Omega_{m}}~~V^4_{A}~~\frac{8}{3}(n+3)\frac{k_N}{\alpha}
\left( \frac{k}{k_N}\right)^{7+2n}_{k=\frac{l}{\tau_0-\tau_{ri}}}
\end{equation}
We define $\Delta T = T_0 [l(l+1)C_l/2\pi]^{1/2}$ as a measure of the
temperature anisotropy, where $T_0=2.73~K$ is the present CMB temperature. 
Taking the square root of the above expression and  
using the values of $V_{A}$, $k_N$, $\Omega_{m}$ and 
$z_{ri}$ mentioned earlier we get,
\begin{equation}
\Delta T=0.52~N_2~B^2_{-9}(n+3)^{1/2}f^{-(n+3)}(\frac{l}{9854})^{3.5+n}~\mu K
\end{equation}
(Note that $k_N (\tau_o-\tau_{ri}) = f 9854$).
For a field of $3$ nano-gauss  ($B_{-9}=3$), a nearly scale invariant 
spectrum for the magnetic field ($n=-2.9$) and for an optical depth of 
$\sim 0.17$ or $N_2 \sim .17$, we get,
\begin{equation}
\Delta T = 0.25\left( \frac{l}{9854} \right)^{0.6}     \label{large_l}
\end{equation}

A similar analysis in the opposite limit
($g_2$ more sharply peaked than $j_l$ and hence 
$l \ll (\tau_0 - \tau_{ri})/\alpha \sim 12$) gives
\begin{equation}
\Delta T=5.3 \times N_2B^2_{-9}(n+3)^{1/2} \left(\frac{l}{9854}\right)^{4+n}~\mu K
\end{equation}
For a field of $3$ nano-gauss  ($B_{-9}=3$), a nearly scale invariant spectrum for 
the magnetic field ($n=-2.9$) and for an optical depth of $0.17$
or $N_2 \sim .17$, we now get,
\begin{equation}
\Delta T = 2.56\left(\frac{l}{9854}\right)^{1.1}~\mu K   \label{small_l}
\end{equation}
For example at $l \sim 10$ one predicts $ 1.3 \times 10^{-3} \mu$K, which
is very small compared to other signals expected at these
low $l$ values.

\section{Numerical Results} \label{numres}

While the calculations in the semi-analytic approximation in the last section 
provide us with rough estimates, to make concrete predictions we need 
to numerically compute the temperature anisotropy. In this section we 
give the results of the numerical evaluation. Specifically, we numerically
evaluate the integral in equation (\ref{Cl}). 
As discussed earlier, the visibility
function appearing in this integral has a dominant contribution at the epoch
of standard recombination and another at the epoch of re-ionization.
The contribution at the epoch of re-ionization we have denoted by $g_2$ and 
we are interested in the additional contribution to $C_l$ resulting from $g_2$.
Unlike the semi-analytic case, 
(where we approximated $g_2$ as an exponential decay
for epochs later than the re-ionization epoch) we use the  
form given in equation (\ref{g}) with the number the number density
of the electrons as given in equation (\ref{ne_theta_f}). 
The resulting expression for $g_2$ is,
\begin{equation}
g_2(\tau_0,\tau)=\frac{3\Omega_{m0}H^2_0\sigma_T}{8\pi G m_b}
\frac{\tau^4_0}{\tau^4}
\Theta(\tau-\tau_{ri})
\exp \left[-\frac{3\Omega_{m0}H^2_0\sigma_T}{8\pi G m_b}
\int^{\tau_0}_{\tau_{ri}}d{\tau'}~(\tau_0/\tau')^4 \right]
\end{equation}  
Further, as the universe is believed to have been matter 
dominated after $\tau=\tau_{ri}$, the redshift and the 
conformal time are related by, $1+z=(\tau_0/\tau)^2$. We can neglect the 
accelerated phase of the universe, as this phase 
is believed to have set in at a low redshift ($z \sim 1$) 
by which time, the visibility function would have 
decayed sufficiently. Hence, this 
will not introduce any significant error in our 
computation. With these simplifying 
assumptions, we have computed $\Delta T$
by evaluating the $\tau$ and $k$ integrals numerically in Eq (\ref{Cl}). While 
evaluating this we have retained the analytical expression for $I(k)$ given in 
equation (\ref{I2-b}). For $B_0 \sim 3$ nano Gauss and a nearly scale invariant 
spectrum ($n=-2.9$) we find $\Delta T = 0.33 \mu K$ for $l=10000$. For higher
$n$ the fluctuations are larger. The results of the numerical 
calculation are shown in Figure 1,
for the magnetic spectral index $n=-2.9$, $-2.8$ and $-2.7$. 
We can compare this with the semi-analytic results 
from equation (\ref{large_l}).
For $n=-2.9$, our semi-analytic calculations give $\Delta T=0.25$ 
for $l=10000$. 
So we see that although the semi-analytic calculation 
marginally underestimates the
temperature anisotropy, we do get the correct order. 
This is the observed trend for all $n$.
For $n=-2.8$ and $n=-2.7$, $\Delta T$ turns out to be $0.47$ $\mu$ K
and $0.58$ $\mu$ K respectively for $l \sim 10000$.
Indeed both the amplitude of $\Delta T$ and its $l$ dependence,
computed using the semi-analytic calculation, agrees
reasonably well with the more exact numerical integration.

\section{Discussion and Conclusions}

We have investigated the additional CMB temperature anisotropies 
that are generated by tangled cosmological magnetic fields in a 
Universe that underwent a relatively early re-ionization. We are motivated
by WMAP results, which indicate that the universe could have been
re-ionized as early as $z \sim 15$. 
We have focused for the present on rotational velocity perturbations, which
can be substained only by cosmological magnetic fields. 
These modes also suffer a milder damping due to
the finite thickness of the re-ionized last scattering surface,
than compressional modes.
Our results supplement earlier work obtained in the context of a 
Universe that does not undergo re-ionization.

We find that a nearly scale-invariant spectrum of tangled 
magnetic fields (with $n=-2.9$ to $n=-2.7$), 
which redshift to a present day value of about 
$3$ nano Gauss  can produce anisotropies at the level of about 
$0.3 \mu$K to $0.5 \mu$K for $l \sim 10000$. 
Even larger signals would obtain if we were to
consider models with larger $n$.
We have simply stopped at this large $l$ because we cannot
use linear theory at present to calculate the expected signals
for $l$ larger than $k_N(\tau_0 -\tau_{ri}) \sim f 9854$.
It is also interesting to note that, at these large $l$,
the above signals are comparable to primary signals
due to tangled fields, arising from the usual recombination 
epoch \cite{compare}.  CMBR anisotropy experiments which
probe such very small angular scales can thus be useful 
to measure or impose bounds on the magnitude as well as the
spectral index of these magnetic fields.
We have also considered here only the simplest first order effects
due to re-ionization. It would be of interest also to estimate the effects
of an inhomogeneous re-ionization on the CMB anisotropy
from vector modes.

\appendix
\section{}

We need to calculate the rotational 
velocity field of the baryons in the post-recombination era, in particular 
around the epoch of re-ionization. We shall follow the formalism developed by 
Wasserman \cite{wasserman}. Here we summarize briefly the essential features 
of the derivation based on Wasserman's paper. 

The Fourier space, linearized Euler equation for the rotational 
component of the baryon velocity in the post recombination era is given by,
\begin{equation}
\frac{\partial {v_i}}{\partial t} + \frac{\dot{a}}{a}{v_i} 
= \frac{P_{ij} F_j}{ 4 \pi \rho_b(t) a^5}
\label{rhob} 
\end{equation}
Here $v_i(k,t)$ is the Fourier component of the rotational
velocity perturbation, $F_j$ is the Fourier component of
the vector $ [({\bf \nabla} \times {\bf B}_0) \times {\bf B}_0]_j$ and
$P_{ij} = \delta_{ij} - k_ik_j/k^2$ as usual
projects out its rotational component.
Note that $V(k,t)$ is the magnitude of the vector $v_i(k,t)$, that is
$\vert V \vert^2 = v_i v_i^*$.
We have also assumed here, (as mentioned in the text)
that on large enough scales, larger than $k_N^{-1} \sim {\rm Mpc}$,
the velocities are so small that the
magnetic field does not get significantly distorted, but simply
redshifts away as ${\bf B}({\bf x},t) = {\bf B}_0/a^2$.

The above equation can be solved in a straightforward fashion
to obtain the rotational component of the velocity field. 
Note that in the post-recombination era, we can neglect 
the radiation density compared to the matter density.
Further, the dark energy component dominated over matter at late 
times, {\it i.e.}, at redshift less than unity, whereas the 
results from WMAP can be 
interpreted to mean that 
the Universe underwent a re-ionization phase at an earlier time. Thus 
from the epoch of recombination ($t_{rec}$), till the epoch of re-ionization 
($t_{ri}$), pressureless matter was the dominant component of the Universe. 
Hence, for $t_{rec} \le t \le t_{ri}$ we can take 
$a(t) = a_{rec} (t/t_{rec})^{2/3} = (1+z)^{-1}$.
The solution for Eq.~\ref{rhob} is then given by
\begin{equation}
v_i=\frac{3t_{rec} (P_{ij} F_j) }{4 \pi \rho_{b0} a^2_{rec}}
\left[  \left(\frac{t_{rec}}{t}\right)^{1/3} 
-\left(\frac{t_{rec}}{t}\right)^{2/3} \right].
\label{vi}
\end{equation}
Here $\rho_{b0}$ is the present day Baryon density, and
we have assumed that the rotational velocity was negligible
at recombination. (Any small rotational velocity at recombination
contributes to a faster decaying term in the above solution than the
term we will retain below).
Since $a(t) \propto t^{2/3}$, $H_{rec}=2/(3t_{rec})$.
Also we have from Einstein equation 
$H^2_{rec}=8\pi G\rho_{rec}/3
=H^2_0~{\Omega_{m0}}~(1+z_{rec})^3$.
Thus 
\begin{equation}
t_{rec}=\frac{2}{3H_0}\frac{(1+z_{rec})^{-3/2}}{\sqrt{\Omega_{m0}}}
\end{equation}
Substituting this expression for $t_{rec}$ in Eq. ~\ref{vi}, and
noting that $a(t) = 1/(1+z)$, we have,
\begin{equation}
v_i=\frac{2}{H_0}\frac{1}{\sqrt{\Omega_{m0}}}  \frac{P_{ij} F_j}
{4\pi \rho_{b0}} \times (1+z)^{1/2}
\label{vifin}
\end{equation}
For a power spectrum of ${\bf B}_0$ as given in the text,
a tedious but straightforward computation gives the power spectrum
of the rotational component of the Lorentz force. We have
\begin{equation}
\frac{k^3}{2\pi^2}\frac{
{\langle}\mid P_{ij}F_j P_{il}F^*_l \mid{\rangle}}{4\pi\rho_b0} = 
\frac{k^2V_A^4}{8} I^2(k)
\end{equation}
where $I(k)$ is a mode coupling integral whose explicit form is given in
\cite{SB98a,ks_trs_jdb}. 
The magnetic field enters through the baryon Alfven velocity $V_{A}$.
The power-spectrum of the $V$ (or the magnitude of $v_i$) 
required in the text is then given by
\begin{equation}
\frac{k^3}{2\pi^2}{\langle}\mid V(k,z)  \mid^2{\rangle} = 
\left(\frac{2}{H_0}\right)^2\frac{1+z}{\Omega_{m0}}\left[\frac{kV^2_{A}}
{\sqrt{8}}I(k)\right]^2
\end{equation}

\begin{figure}
\caption{The figure shows the temperature anisotropy due to
vector type modes induced by tangled 
cosmological magnetic fields that are of a strength of 3 nano Gauss today. 
The magnetic power spectrum is assumed to be a power-law 
characterized by an index $n$. 
The figure shows the results for $n=-2.9$ (solid line), $-2.8$ (dotted line)
and $-2.7$ (dashed line).}
\centering
\begin{picture}(500,500)
\psfig{figure=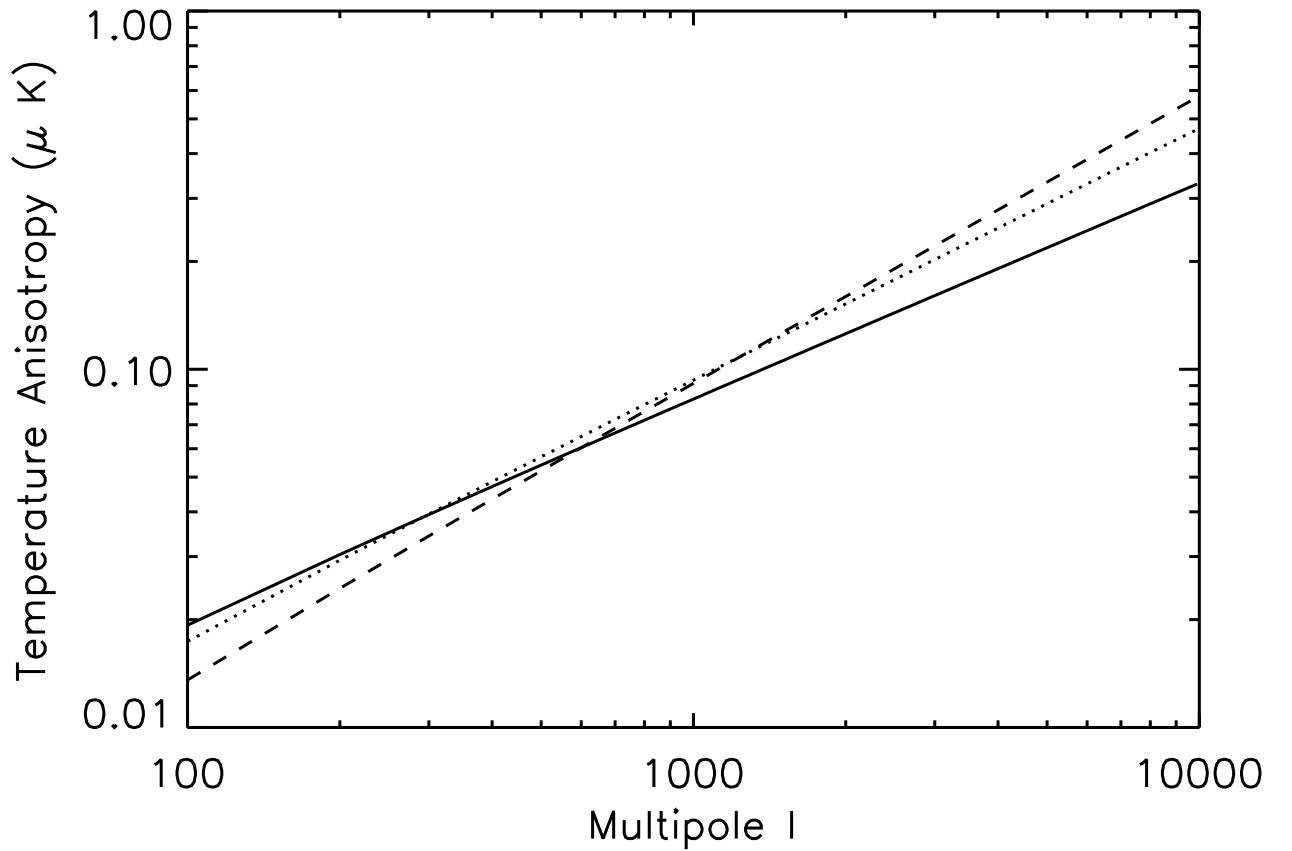}
\end{picture}

\end{figure}

{\it Acknowledgments:} TRS thanks IUCAA for the support provided
through the Associateship Program and the facilites at
the IUCAA Reference Centre at Delhi University.

\end{document}